\documentclass[twocolumn,8pt,nobalance]{article}
\usepackage{lmodern}
\usepackage{amssymb,amsmath}
\usepackage[usenames,dvipsnames]{color}
\usepackage{longtable,booktabs}
\usepackage{graphicx,grffile}
\usepackage[left=.5in, right=.5in, top=.75in,bottom=.75in]{geometry}
\usepackage{authblk}
\usepackage{tabularx}
\usepackage{stfloats}
\usepackage{txfonts}
\usepackage{xspace}
\usepackage{algorithm}
\usepackage{subcaption}
\usepackage{listings}
\usepackage{booktabs}
\usepackage{enumitem}
\usepackage{makecell}
\usepackage{titlesec}
\usepackage{placeins} 
\usepackage{xr}
\providecommand{\keywords}[1]{\textbf{\textit{Keywords:\xspace}} #1}

\usepackage{fancyhdr}
\newcommand{\OPENRBC}[0]{\textbf{\sffamily OpenRBC}\xspace}
\newcommand{\OpenRBC}[0]{OpenRBC\xspace}
\titlespacing\section      {0pt}{3pt plus 1pt minus 1pt}{3pt plus 1pt minus 1pt}
\titlespacing\subsection   {0pt}{3pt plus 1pt minus 1pt}{3pt plus 1pt minus 1pt}
\titlespacing\subsubsection{0pt}{3pt plus 1pt minus 1pt}{3pt plus 1pt minus 1pt}

\let\OLDthebibliography\thebibliography
\renewcommand\thebibliography[1]{
  \OLDthebibliography{#1}
  \setlength{\parskip}{0pt}
  \setlength{\itemsep}{0pt plus 0.3ex}
}

\titleformat{\section}{\normalfont\fontsize{14}{16}\bfseries}{\thesection}{1em}{}

\title{\vspace{-1cm}\OPENRBC: A Fast Simulator of Red Blood Cells at Protein Resolution}
\author[*1]{Yu-Hang Tang}
\author[*1]{Lu Lu}
\author[1]{He Li}
\author[2]{Constantinos Evangelinos}
\author[2]{Leopold Grinberg}
\author[2]{Vipin Sachdeva}
\author[1]{George Em Karniadakis}
\affil[1]{Division of Applied Mathematics, Brown University, Rhode Island, USA 02912}
\affil[2]{International Business Machines Corporation}
\affil[*]{Equally credited}
\date{}

\definecolor{mygreen}{rgb}{0,0.6,0}
\definecolor{mygray}{rgb}{0.5,0.5,0.5}
\definecolor{mymauve}{rgb}{0.58,0,0.82}

\begin{document}
\maketitle

\abstract{We present \OpenRBC\footnote{The source code is available at {\color[rgb]{0,0.3,0.8} \texttt{https://github.com/yhtang/OpenRBC} }.}, a coarse-grained molecular dynamics code,
which is capable of performing an unprecedented \emph{in silico} experiment --- simulating an
entire mammal red blood cell lipid bilayer and cytoskeleton as modeled by 4 million
mesoscopic particles --- using a single shared memory commodity workstation.
To achieve this, we invented an adaptive spatial-searching algorithm to accelerate the computation
of short-range pairwise interactions in an extremely sparse 3D space. The
algorithm is based on a Voronoi partitioning of the point cloud of coarse-grained particles, and
is continuously updated over the course of the simulation. The algorithm
enables the construction of the key spatial
searching data structure in our code, \textit{i.e.} a lattice-free cell list, with a time and space cost linearly proportional
to the number of particles in the system. The position and shape of the cells also adapt automatically to the local density and curvature.
The code implements OpenMP
parallelization and scales to hundreds of hardware threads.
It outperforms a legacy simulator by almost an order of magnitude in time-to-solution and more than 40 times
in problem size, thus providing a new platform for probing the biomechanics of red blood cells.}

\keywords{coarse-grained molecular dynamics, bilayer, cytoskeleton, membrane fluctuation, vesiculation, high-performance computing}

\section*{Introduction}\label{introduction}

The red blood cell (RBC) is one of the simplest, yet most important
cells in the circulatory system due to their indispensable role in oxygen transport. An
average RBC assumes a biconcave shape with a diameter of 8 \(\mu\)m and
a thickness of 2 \(\mu\)m. Without any intracellular organelles, it is supported by a cytoskeleton of a triangular spectrin network
anchored by junctions on the inner side of the membrane. Therefore, the mechanical properties of an RBC can be strongly influenced by molecular level structural details that alter the cytoskeleton and lipid bilayer properties.

Both continuum models~\cite{evans1974bending, feng2006finite, powers2002fluid, helfrich1973elastic} and particle-based models~\cite{feller2000molecular, saiz2002towards, tieleman1997computer, tu1998constant}
have been developed with the aim to help uncover the correlation between RBC membrane structure and property. Continuum models are computationally efficient, but require \textit{a priori} knowledge of cellular mechanical properties such as bending and shear modulus.
Particle models are useful for extracting RBC properties from low-level descriptions of the membrane structure and defects.
However, it is computational demanding, if not prohibitive, to simulate the large number of particles required for modeling the membrane of an entire RBC.
To the best of our knowledge, a bottom-up simulation of the RBC membrane at the cellular scale using particle methods remains absent.

Recently, a two-component coarse-grained molecular dynamics (CGMD) RBC membrane model which explicitly accounts for both the cytoskeleton and the lipid bilayer was proposed~\cite{li2014erythrocyte}.
The model could potentially be used for particle-based whole-cell RBC modeling 
because its coarse-grained nature can drastically reduce computational workload while still preserving necessary details from the molecular level.
However, due to the orders of magnitude difference in the length scale between a cell and a single protein, a total of 4 million particles are still needed to represent an entire RBC.
In addition, the implicit treatment of the plasma in this model eliminates the overhead for tracking the solvent particles,
but also exposes a notable spatial density heterogeneity because all CG particles are exclusively located on the surface of a biconcave
shell. The inside and outside of the RBC are empty. This density imbalance imposes a serious challenge on the efficient evaluation of the pairwise force using conventional algorithms and data structures, such as the cell list and the Verlet list,
which typically assumes a uniform spatial density and a bounded rectilinear simulation box.

\begin{figure*}
  \begin{minipage}[c]{0.67\textwidth}
    \includegraphics[width=\textwidth]{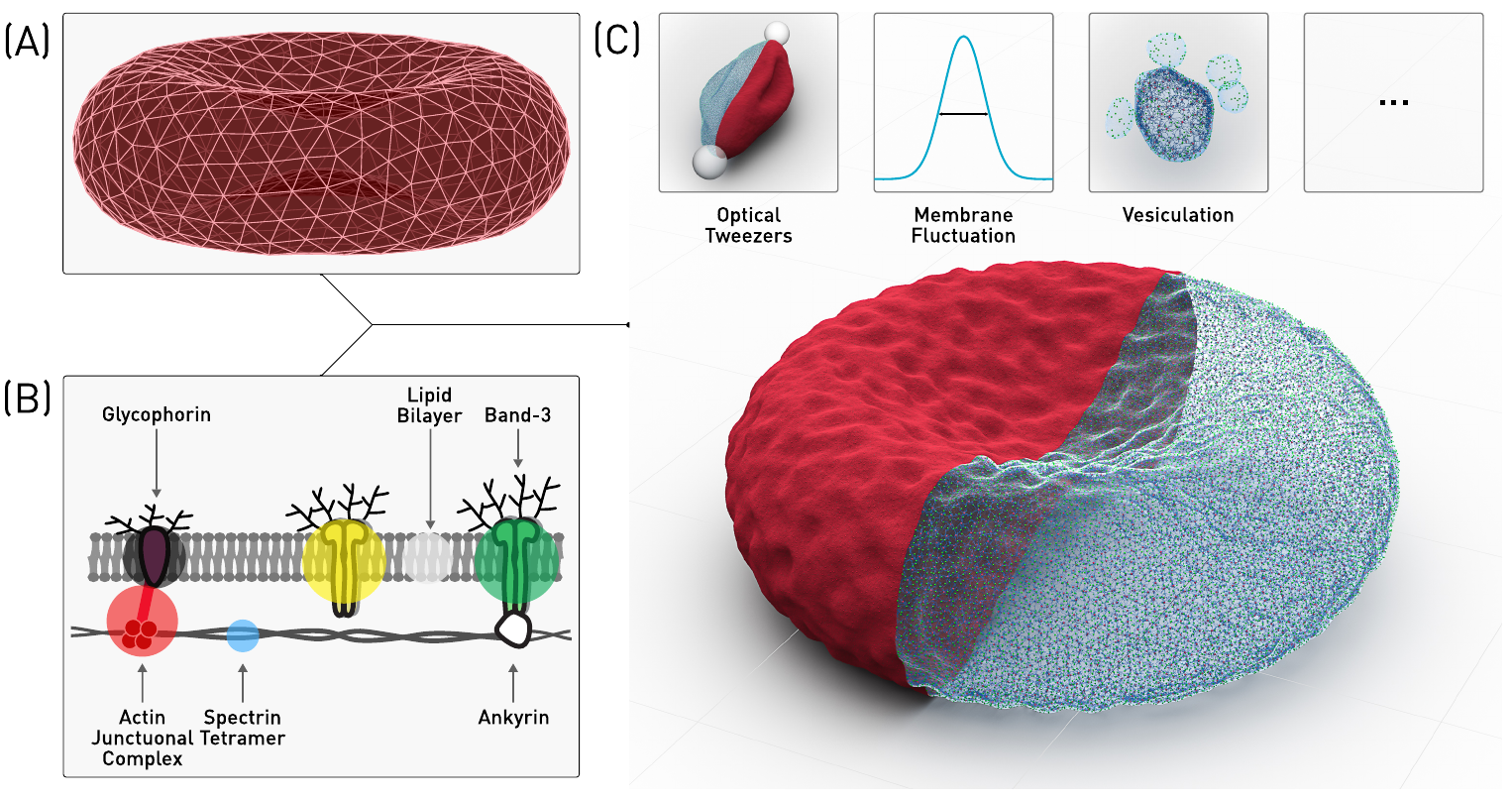}
  \end{minipage}\hfill
  \begin{minipage}[c]{0.3\textwidth}
    \caption{
      \textbf{(A)} A canonical hexagonal triangular mesh of a biconcave surface representing the cytoskeleton network \label{fig:overview}
      is used together with \textbf{(B)} the two-component CGMD RBC membrane model to reconstruct \textbf{(C)} a full-scale virtual RBC which allows
      for a wide range of computational experiments.\label{fig:cgmd-rbc-scheme}
    }
  \end{minipage}
\end{figure*}

In this paper we present \OpenRBC, a new software tailored for simulations of an entire RBC using the two-component CGMD model on multicore CPUs.
As illustrated in Figure~\ref{fig:overview}, the simulator can take as input a 
triangular mesh of the cytoskeleton of a RBC and reconstruct a
CGMD model at protein resolution with explicit representations
of both the cytoskeleton and the lipid bilayer.
This type of whole cell simulation of RBCs
can thus realize an array of \emph{in silico}
measurements and explorations of:
\begin{itemize}[label=$\cdot$,leftmargin=*]
  \setlength\itemsep{-0.5em}
  \item RBC shear and bending modulus,
  \item membrane loss through vesiculation in spherocytosis and elliptocytosis~\cite{li2015vesiculation},
  \item anormalous diffusion of membrane proteins~\cite{li2016modeling},
  \item interaction between sickle hemoglobin fibers and RBC membrane in sickle cell disease~\cite{lei2012predicting,li2016patient},
  \item uncoupling between the lipid bilayer and cytoskeleton~\cite{peng2013lipid},
  \item adenosine triphosphate (ATP) release due to deformation~\cite{sprague1998deformation},
  \item nitric oxide (NO) modulated mechanical property change~\cite{wood2013circulating},
  \item cellular uptake of elastic nanoparticles~\cite{zhao2011interaction}.
\end{itemize}

%

\section*{Software Overview}\label{simulator-design}

\OpenRBC is written in C++ using features from the C++11 standard.
To maximize portability and allow easy integration into other software systems\cite{tang2015multiscale}, the project is organized as a header-only library with no external dependencies.
The software implements SIMD vectorization \cite{abraham2015gromacs} and OpenMP shared memory parallelization, and was specifically optimized
toward making efficient use of large numbers of simultaneous hardware threads. 

\begin{figure}[t!]
	\centering
	\includegraphics[width=\columnwidth]{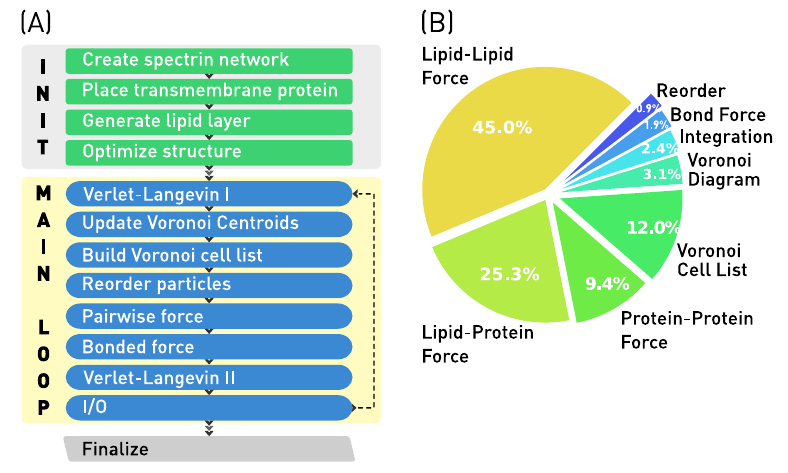}
	\caption{The A) flow chart and B) typical wall time distribution of \OpenRBC.\label{fig:design}}
\end{figure}

As shown in Figure~\ref{fig:design}A, the main body of the simulator is a time stepping loop, where the force and torque acting on each
particle is solved for and used for the iterative updating of the position and orientation according to the equation of motion. The time distribution of each task in a typical simulation is given in
Figure~\ref{fig:design}B. The majority of time is spent in force evaluation, which is compute-bound.
This makes the code highly efficient in utilizing the high thread count of modern CPUs with the shared memory programming paradigm.

\section*{Initial structure generation}

As shown in Figure~\ref{fig:cgmd-rbc-scheme}, a two-component CGMD RBC system can be generated from a triangular mesh which resembles the biconcave shape of
a RBC at equilibrium. Note that the geometry may be alternatively sourced from experimental data using techniques
such as optical image reconstruction because the algorithm itself is general enough to adapt to an arbitrary triangular mesh.
This feature can be useful for simulating RBCs with morphological anomalies.
Actin and glycophorin protein particles are placed on the vertices of the mesh, while spectrin and immobile band-3
particles are generated along the edges. The band-3--spectrin connections and actin--spectrin connections can be modified to simulate RBCs with structural defects.
Lipid and mobile band-3 particles are randomly placed on each triangular face by uniformly sampling each triangle
defined by the three vertices~\cite{osada2002shape}. A minimum inter-particle distance is enforced to prevent clutter between protein and lipid particles.
The system is then optimized using a velocity quenching algorithm to remove collision between the particles.

\section*{Spatial Searching Algorithm}\label{spatial-search-algorithm}

\begin{figure}
\centering
\includegraphics[width=\columnwidth]{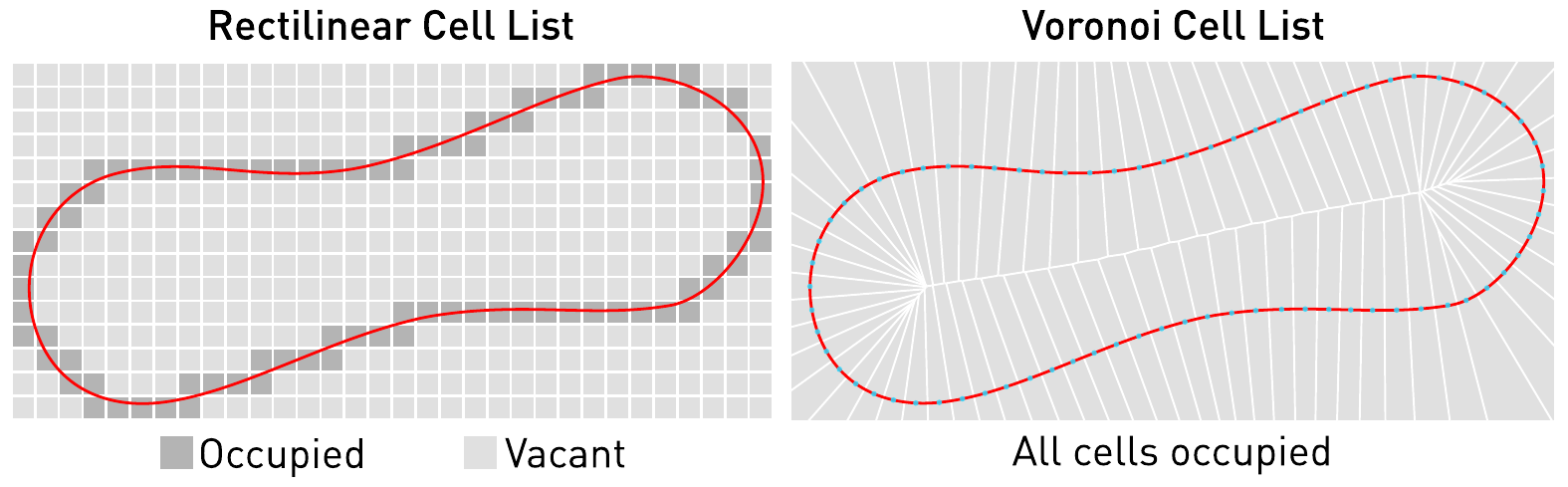}
\caption{\textbf{Left:} Only cells in dark gray are populated by CG particles in a cell list on a rectilinear lattice. This results in a waste of storage and memory bandwidth. \textbf{Right:} All cells are evenly populated by CG particles in a cell list based on the Voronoi diagram generated from centroids located on the RBC membrane. \label{fig:cell-list-vacant}}
\end{figure}

Pairwise force evaluation accounts for more than 70\% of the computation time in \OpenRBC as well as other molecular dynamics softwares\cite{tang2014accelerating,Rossinelli2015GB15}.
To efficiently simulate the reconstructed RBC model, we invented a lattice-free spatial partitioning algorithm that is inspired by the concept of Voronoi diagram.
The algorith, at the high level, can be described as
\begin{lstlisting}
1. Group particles into a number of `adaptive' clusters
2. Compute interactions between neighboring clusters
3. update cluster composition after particle movement
4. repeat from step 2
\end{lstlisting}
As illustrated in Figure~\ref{fig:cell-list-vacant}, the algorithm adaptively partitions a particle system into a number of Voronoi cells that are approximately equally populated.
In contrast, a lattice-based cell list leaves many cells vacant due to the density heterogeneity.
Thus, the algorithm can provide very good performance in partitioning the system, maintaining data locality and searhcing for pairwise neighbors in a sparse 3D space.
It is implemented in our software using a \textit{k}-means clustering algorithm, which is, in turn, enabled by a highly optimized implementation of the \textit{k}-d tree searching algorithm, as explained below.

\begin{figure}
\centering
\includegraphics[width=\columnwidth]{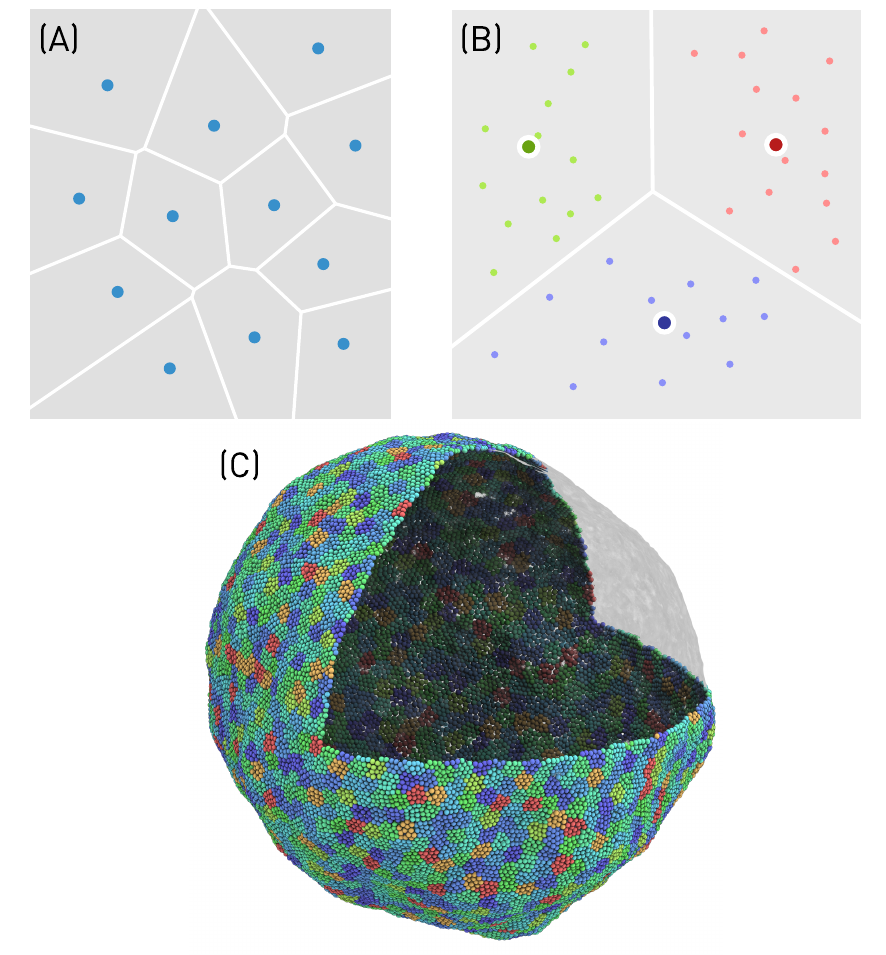}
\caption{(A) A Voronoi partitioning of a square as generated by centroids marked by the blue dots. (B) A \textit{k}-means (\textit{k}=3) clustering of a number of points on a 2D plane.
(C) A vesicle of 32,673 CG particles. 
\label{fig:voronoi-kmeans}}
\end{figure}

A Voronoi tessellation \cite{edelsbrunner2014voronoi} is a partitioning of a n-dimensional space into
regions based on distance to a set of points called the centroids. Each point
in the space is attributed to the closest centroid (usually in the L2
norm sense). An example of a Voronoi Diagram generated by 12 centroids on a 2D rectangle is given in Figure~\ref{fig:voronoi-kmeans}A.

The \textit{k}-means clustering \cite{hartigan1979algorithm} is a method of data partitioning that aims to
divide a given set of n vectors into k clusters in which each vector
belongs to the cluster whose center is closest to it. The result is a
partition of the vector space into a Voronoi tessellation generated by
the cluster centers as shown in Figure~\ref{fig:voronoi-kmeans}B. Searching for the optimal clustering which minimizes
the within-cluster sum of square distance is NP-hard, but efficient iterative heuristics based on \textit{e.g.}
the expectation-maximization algorithm \cite{dempster1977maximum} can be used
to quickly find a local minimum.

\begin{figure}
\centering
\includegraphics[width=\columnwidth]{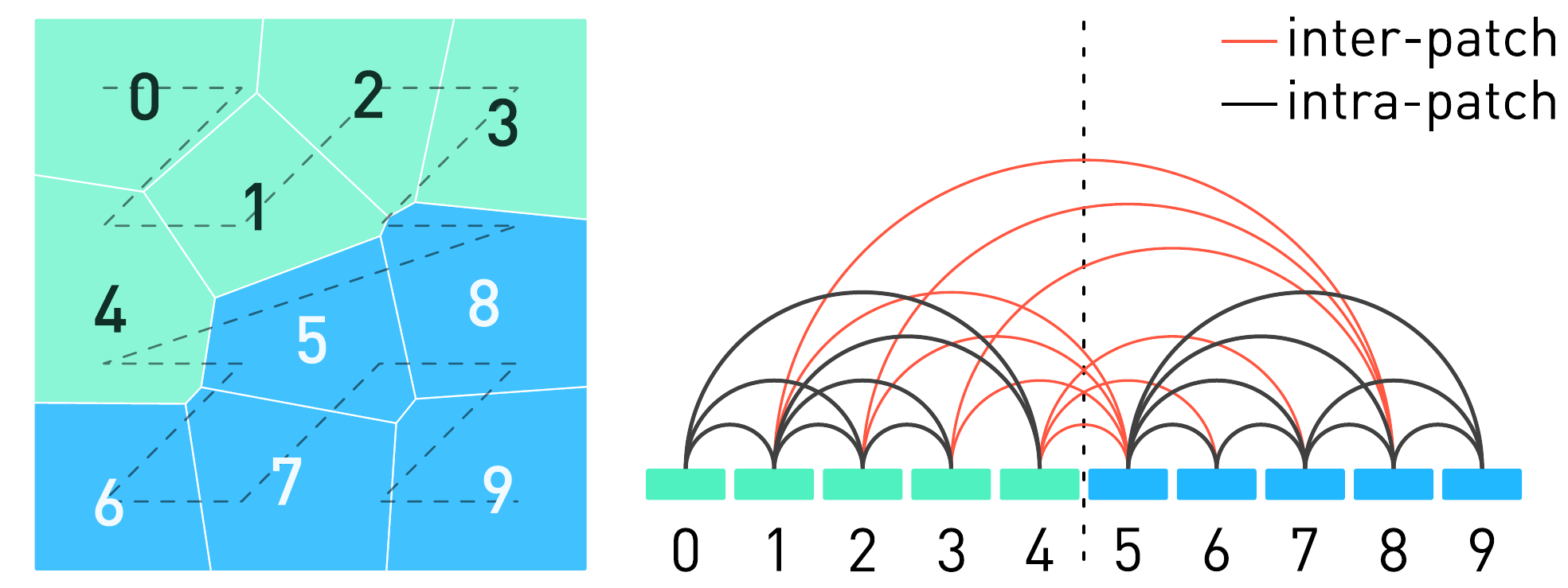}
\caption{Thanks to the spatial locality ensured by reordering particles along the Morton curve (dashed line), we can simply divide the cells between two threads by their index into two patches each containing five consective Voronoi cells. The force between cells from the same patch is computed only once using Newton's 3rd law, while the force between cells from different patches is computed twice on each side.
\label{fig:patch-force}}
\end{figure}

A \textit{k}-d tree is a spatial partitioning data structure for organizing points in a \textit{k}-dimensional space~\cite{bentley1975multidimensional}.
It is essentially a binary tree that recursively bisects the points owned by each node into two disjoint sets as separated by a axial-parallel hyperplane.
It can be used for the efficient searching of the nearest neighbors of a given query point
in $O(\log N)$ time, where $N$ is the total number of points, by pruning out a large portion of the search space using cheap overlap checking between bounding boxes.

The \textit{k}-means/Voronoi partitioning of a point cloud adapts automatically to the local density and curvature of the points.
As such, we exploit this property to create a generalization of the cell list algorithm using the
Voronoi diagram. The algorithm can be described
as a two-step procedure: 1) clustering all the particles in the system using \textit{k}-means,
followed by an online Expectation-Maximization algorithm that
continuously updates the system's Voronoi cells centroid location and
particle ownership; 2) sorting the centroids and particles with a two-level data reordering scheme
, where we first order the Voronoi centroids along a space
filling curve (a Morton curve, specifically) and then reorder the particles
according to the Voronoi cell that they belong to. The pseudocode for the algorithm can be found in SI Algorithm~\ref{SI-alg:voronoi-cell-list}.
The reordering step in updating the Voronoi cells ensures that neighboring particles in the
physical space are also statistically close to each other in the
physical space. This locality can speed up the \textit{k}-d tree nearest-neighbor search by
using the closest centroid of the last particle as the
initial guess for the current particle. The heuristic helps to further prune out most of the \textit{k}-d tree search space and essentially reduces the
complexity of a nearest-neighbor query from \(O(\log N)\) to \(O(1)\).
In practice, this brings about 100 times acceleration when searching through 200,000
centroids. As shown by Figure~\ref{fig:voronoi-kmeans}C, the Voronoi cells
generated from a \textit{k}-means clustering of the CG particles are uniformly distributed on the surface of the
lipid membrane.

\section*{Force Evaluation}\label{force-evaluation}

Lipid particles accounts for 80\% of the population in the whole-cell CGMD system.
The Voronoi cells can be used directly for efficient pairwise
force computation between lipids with a quad loop
that ranges over all Voronoi cell \(v_i\), all neighboring cells \(v_j\) of \(v_i\), all particles in
\(v_i\), and all particles in \(v_j\) as shown in SI Algorithm~\ref{SI-alg:pairwise}.
Since the cytoskeleton of a healthy RBC is always attached to the lipid bilayer, its protein particles are also distributed following
the local curvature of the lipid particles. This means that we can reuse the Voronoi cells of the lipid particles, but with 
a wider searching cutoff, to compute both the lipid-protein and protein-protein pairwise interactions.
For diseased RBCs with a fully or partially detached cytoskeleton, a separate set of Voronoi cells 
can be set up for the cytoskeleton proteins to compute the force.
A list of bonds between proteins is maintained and
used for computing the forces between proteins that are physically
linked to each other.

A commonly used technique in serial programs to speed up the force
computation is to take advantage of the Newton's 3rd law of action and
reaction. Thus, the force between each pair of interacting particles is
only computed once and added to both particles. However, this generates
a race condition in a parallel context because two threads may
end up simultaneously computing the force on a particle shared by two or more pairwise interactions.

Our solution takes advantage of the strong spatial locality of the
particles as maintained by the two-level reordering algorithm, and decomposes the
workload both spatially and linearly-in-memory into patches by splitting the linear range of cells
indices among OpenMP threads.
Each thread will be calculating the forces acting on the particles within its own
patch. As shown in Figure~\ref{fig:patch-force}, force accumulation without triggering racing condition can be
realized by only exploiting the Newton's 3rd law on pairwise
interactions where both Voronoi cells belong to a thread's own patch.
Interactions involving a pair of particles from different patches are calculated twice
(once for each particle) by each thread. The strong particle
locality minimizes the shared contour length between two patches and
hence also minimizes the amount of inter-patch interactions.

%


\section*{Validation and Benchmark}\label{benchmark}

In this section  we present validation of our software by comparing simulation and experimental data.
We also compare the program performance against that of a legacy CGMD RBC simulator used in Ref~\cite{li2014erythrocyte}.
The legacy simulator, which performs reasonably well for a small number of particles in a periodic rectangular box, was written in C and parallelized with the message passing interface (MPI) using a rectilinear domain decomposition scheme and a distributed memory model.
Two computer systems were used in the benchmark each equipped with a different mainstream CPU microarchitecture, \textit{i.e.} the AMD Piledriver, and the IBM Power8~\cite{starke2015cache}.
The specification of the machines are given in Table~\ref{table:computer-specs}.

\begin{table*}[htbp!]
\centering
\caption{A summary of capability and design highlights of \OpenRBC and the specifications of the computer systems used in benchmark.\label{table:computer-specs}}

\begin{tabularx}{\textwidth}{llllXlllll}
\toprule
\multicolumn{4}{c}{\textbf{Capability \& Design}}                                       && \multicolumn{5}{l}{\textbf{Performance - time steps / day}} \\ \midrule
                                        & \textbf{\OpenRBC}  & \textbf{Legacy}  & \textbf{Improvement} && \textbf{Cores} & \textbf{Particles} & \textbf{\OpenRBC}   & \textbf{Legacy} & \textbf{Speedup} \\
\textbf{Max system size (\#particles)}  & $> 8 \times 10^6$  & $2 \times 10^5$  & $>$ 40 times         && 20             & $8.34 \times 10^6$ & $3.90\times 10^5$ & \textit{-}          & -                \\
\textbf{Line of code}                   & 4,677              & 7,424            & 37\% less            && 4              & $1.88 \times 10^5$ & $3.86\times 10^6$ & $0.42\times 10^6$            & 9.2              \\ \bottomrule
\end{tabularx}

\begin{tabularx}{\textwidth}{lllllllllll}
\toprule
\thead[tl]{CPU}          & \thead{Architecture} & \thead[tl]{Instruction\\Set} & \thead[tl]{Freq.\\(GHz)} & \thead[tl]{Physical\\Cores} & \thead[tl]{Hardware\\Threads} & \thead[tl]{Total\\threads} & \thead[tl]{Last Level\\Cache (MB)} & \thead[tl]{GFLOPS\\(SP)} & \thead[tl]{Achieved FLOPS\\by \OpenRBC} & \thead[tl]{Memory\\Bandwidth} \\
\midrule
IBM POWER 8 `Minsky'  & Power8     & Power  & 3.5 & 10 $\times$ 2 & 8 & 160 & 80 $\times$ 2 & 560.0  & 8.7\% & 230 GB/s \\
AMD Opteron 6378      & Piledriver & x86-64 & 2.4 & 16 $\times$ 4 & 1 & 64  & 16 $\times$ 4 & 614.4  & 4.5\% & 204 GB/s \\
\bottomrule
\end{tabularx}
\end{table*}

\begin{figure}
\centering
\includegraphics[width=\columnwidth]{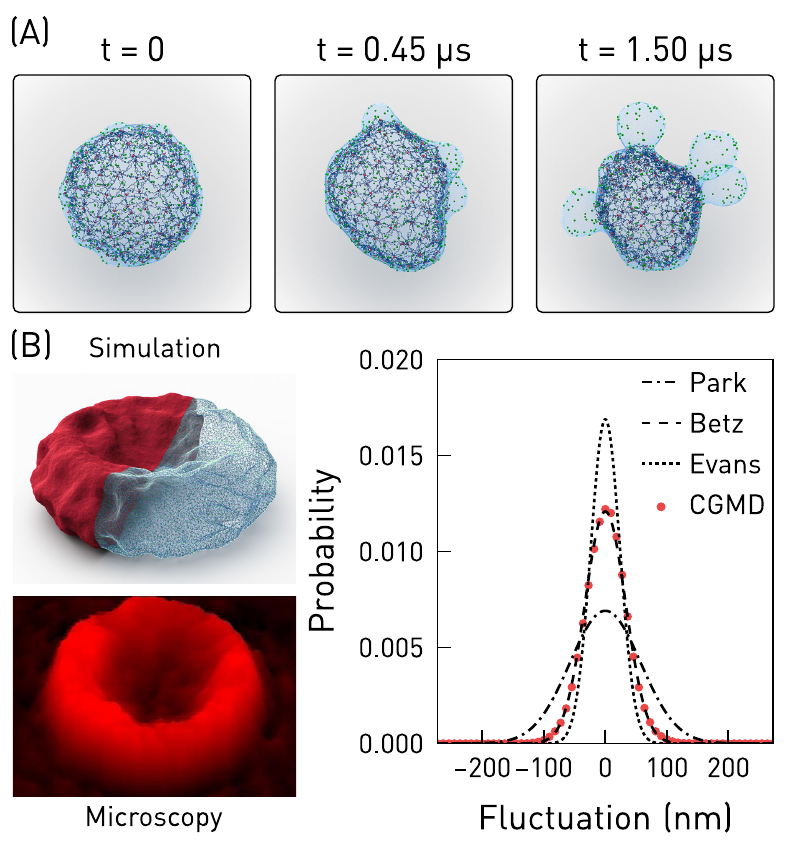}
\caption{(A) The vesiculation procedure of a miniature RBC. (B) The instantaneous fluctuation of a full-size RBC in \OpenRBC compares to that from experiments~\cite{evans2008fluctuations,park2008refractive,betz2009atp}. Microscopy image reprinted with permission from Ref.~\cite{park2008refractive}. \label{fig:validation}}
\end{figure}

To compare performance between \OpenRBC and the legacy simulator, the membrane vesiculation process of a miniaturized RBC-like sphere with a surface area of of $2.8\ \mu m^2$ was simulated.
The evolution of the dynamic process is visualized from the simulation trajectory and shown in Figure~\ref{fig:validation}A. \OpenRBC achieves almost an order of magnitude speedup over the legacy solver in this case on all three computer systems as shown in Table~\ref{table:computer-specs}.

%


Furthermore, \OpenRBC can efficiently simulate an entire RBC modeled by 3.2 million particles and correctly reproduce the fluctuation and stiffness of the membrane as shown in Figure~\ref{fig:validation}B.
The legacy solver was not able to launch the simulation due to memory constraint.
The simulation was carried out by implementing the experimental protocol of Ref.~\cite{park2008refractive} that measures the instantaneous vertical fluctuation $\Delta h(x,y)$ along the upper rim of a fixed RBC. In addition, a harmonic volume constraint is applied to maintain the correct surface-to-volume ratio of the RBC. We measured a membrane root-mean-square displacement of 33.5 nm, while previous experimental observations and simulation results range between 23.6 nm to 58.8 nm~\cite{evans2008fluctuations,park2008refractive,betz2009atp,fedosov2011multiscale}.

\begin{figure}[b!]
\centering
\includegraphics[width=\columnwidth]{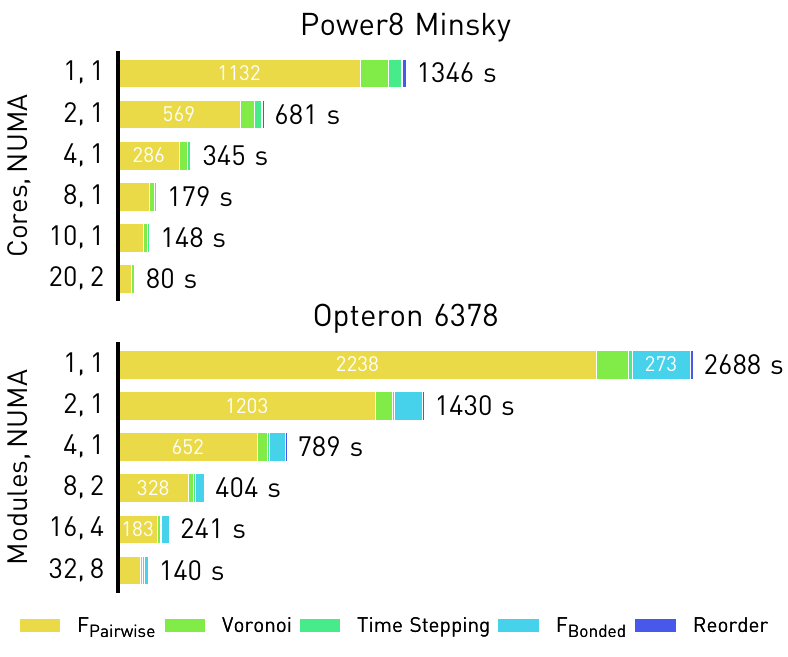}
\caption{Scaling of \OpenRBC across physical cores and NUMA domains when simulating an RBC of 3.2 million particles.\label{fig:scaling}}
\end{figure}

Scaling benchmark for the whole cell simulation on the three computer systems is given in Figure~\ref{fig:scaling}.
It can be seen that compute-bound tasks such as pairwise force evaluation can scale linearly across physical cores. Memory-bound tasks
benefit less from hardware threading as expected, however thanks to thread pinning and a consistent workload decomposition between threads there is no
performance degradation due to side effects such as cache and bandwidth contention.




It is also worth noting that Fu \textit{et al.} recently published an implementation of a related RBC model in LAMMPS~\cite{fu2017lennard},
which can simulate $1.15 \times 10^6$ particles for $10^5$ time steps on 864 CPU cores in 2761 seconds.
However, the use of explicit solvent particles in their model generates difficulty to establish a fair performance comparison between their implementation and \OpenRBC.
Nevertheless, as a rough estimate and assuming perfect scaling, their timing result can be translated into simulating $8.34 \times 10^6$ particles for $0.41 \times 10^5$ time steps per day on 864 cores.
OpenRBC can perform roughly the same amount of time steps on 20 CPU cores.
We do recognized that the explicit solvent model carrys more computational workload,
and that implementing non-rectilinear partitioning schemes may not be straightforward within the current software framework of LAMMPS.
This comparison serves more as a demonstration of the usage of the shared-memory programming paradigm on \textit{fat} compute nodes with large amounts of strong cores and memory.

\section*{Summary}\label{summary}

We presented a from-scratch development of a coarse-grained molecular dynamics software, \OpenRBC, which exhibits exceptional efficiency when simulating
systems of large density discrepancy. This capability is supported by a key algorithm innovation for computing an adaptive partitioning of the particles
using a Voronoi diagram. The program is parallelized with OpenMP and SIMD vector instructions, and implements threading affinity control, consistency loop partitioning, kernel fusion, and
atomics-free pairwise force evaluation to increase the utilization of simultaneous hardware threads and to maximize memory performance across multiple NUMA domains.
The software achieves an order of magnitude speedup in terms of time-to-solution over a legacy simulator, and can handle systems that are almost two orders of magnitude larger in particle count.
The software enables, for the first time ever, simulations of an entire RBC with a resolution down to single proteins, and
opens up the possibility for conducting many \textit{in silico} experiments concerning the RBC cytomechanics and related blood disorders \cite{li2016computational}.
\footnote{The source code is available at {\color[rgb]{0,0.3,0.8} \texttt{https://github.com/yhtang/OpenRBC} }.}

\titleformat{\section}{\normalfont\fontsize{11}{13}\bfseries}{\thesection}{1em}{}

\scriptsize
\section*{Author Contribution}
\noindent YHT designed algorithm. YHT and LL implemented the software. YHT and CE carried out performance benchmark. HL developed the CGMD model. HL and YHT performed validation and verification. YHT, LL and HL wrote the manuscript. LG, CE, and VS provided algorithm consultation and technical support. GK supervised the work.

\section*{Acknowledgment}
\noindent This work was supported by National Institutes of Health (NIH) grants U01HL114476 and U01HL116323 and partially by the Department of Energy (DOE) Collaboratory on Mathematics for Mesoscopic Modeling of Materials (CM4). YHT acknowledges partial financial support from an IBM Ph.D. Scholarship Award. Part of the simulations were carried out at the Oak Ridge Leadership Computing Facility through the Innovative and Novel Computational Impact on Theory and Experiment program at Oak Ridge National Laboratory under project BIP118.

\bibliography{OpenRBC}
\bibliographystyle{bibstyle}

\end{document}